# THE CMS PIXEL DETECTOR: FROM PRODUCTION TO COMMISSIONING

VINCENZO CHIOCHIA[†]

*On behalf of the CMS Collaboration*

*University of Zürich, Physik-Institut, Winterthurerstr. 190*
*Zürich, CH-8057, Switzerland*

The CMS experiment at the LHC includes a hybrid silicon pixel detector for the reconstruction of charged tracks and of the interaction vertices. The detector is made of three barrel layers and two disks at each end of the barrel. Detector modules consist of thin, segmented silicon sensors with highly integrated readout chips connected by the bump bonding technique. In this paper we report on the progress of the detector construction and testing. In addition, first results from the commissioning systems at CERN and PSI are presented.

## 1. Introduction

The CMS experiment, currently under construction at the Large Hadron Collider (LHC) will include a hybrid silicon pixel detector [1] to allow tracking in the region closest to the interaction point. The detector will be a key component for reconstructing interaction vertices and heavy quark decays in a particularly harsh environment, characterized by a high track multiplicity and heavy irradiation.

In this paper we report on the status of the detector production and testing. In addition the results of the first commissioning tests are discussed. The paper is structured as follows: Section 2 describes the mechanical structure of the detector; in Section 3 the production and testing of the detector modules is detailed. The detector assembly and commissioning tests are described in Section 4 for both engineering and production systems. Conclusions and future plans are given in Section 5.

---

[†] email: vincenzo.chiochia@cern.ch

## 2. Mechanical structure

The detector consists of three barrel layers (BPIX) and two disks (FPIX) at each end of the barrel. The innermost barrel layer has a radius of 4.3 cm, while for the second and third layers the radius is 7.2 cm and 11 cm, respectively. The layers are composed of modular detector units. The modules consist of thin segmented silicon sensors with highly integrated readout chips (ROC) connected by the bump bonding technique. Eight modules are screwed on 53 cm long and 0.25 mm thin carbon fibre ladders, which are glued to aluminium cooling pipes. Ladders are mounted on half-shells, constituting the smallest independent unit of the barrel support structure. Three half-shells are mounted together at the two end flanges building up half of the barrel detector (see Figure 1 left). The left and right half barrels are mechanically separated. Each half detector is connected to two supply-tube half cylinders containing the level translator chips as well as analogue- and digital-optohybrid circuits. In addition, each supply tube half cylinder provides the power and cooling lines to the half barrel detector.

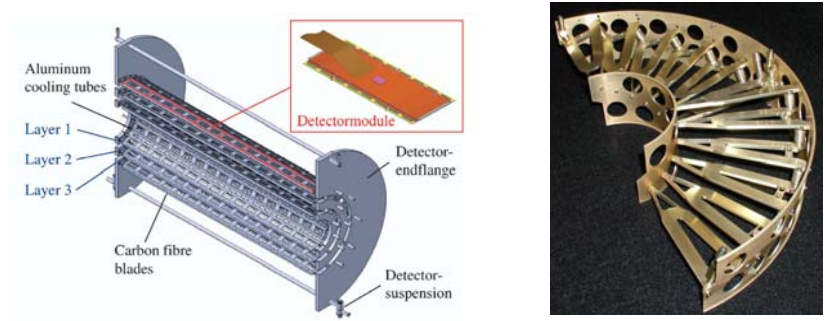

Figure 1: Complete half barrel support structure with three half shells (left). The FPIX half-disk cooling channels mounted in the outer half-ring structure (right). Panels are mounted on both sides of the cooling channels.

The FPIX disks extending from about 6 to 15 cm in radius are placed at $z=\pm 34.5$ cm and $z=\pm 46.5$ cm. Disks are divided into half-disks including 12 "U"-shaped cooling channels disposed in a turbine-like geometry to enhance charge sharing (see Figure 1, right). Each cooling channel has trapezoidal beryllium "panels" attached to each side and supporting the detector modules. Pairs of half-disks are installed in the service half-cylinder containing all the mechanical and electrical infrastructure needed to support, position, cool, power, control and read out the detector.

## 3. Module production and testing

Each BPIX full module (or half module) is composed by two (or one) silicon nitride base strips, 16 (or 8) ROCs [2], a 285 μm thick *n*-in-*n* silicon pixel sensor [3], a High-Density Interconnect (HDI) and a Token Bit Manager (TBM) chip. The number or readout channels per module is 66'560 (or 33'280). Signal and power cables connect each module to the patch panel at each end flange (see Figure 2, left). A complete module has the dimensions 66.6×26.0 mm$^2$, a weight of 2.2 g and generates about 2 W. 672 full modules and 96 half modules are required for BPIX.

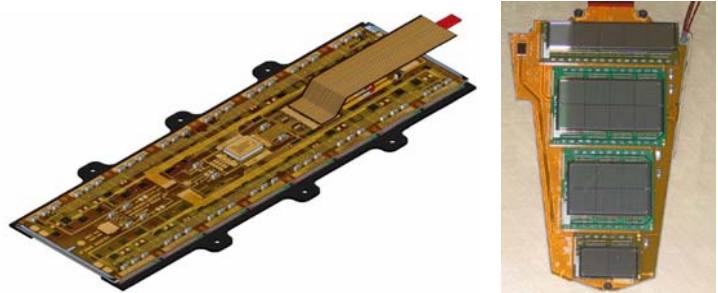

Figure 2: Drawing of a BPIX full module (left). Photograph of a FPIX panel with four plaquettes (right).

The HDI, ROCs and sensors wafers are pre-processed and prepared for module assembly as described in [4]. The process includes the indium deposition on sensors wafers. Only devices with less than 1% of dead or noisy pixels and without pixel masking defects are used in the following assembly steps. Sensors with deposited and re-flown bump bonds are connected to ROCs using a bump-bonding machine developed at PSI. ROCs are tested during bump-bonding to ensure that only good chips are connected to the sensor. The "raw" module is consequently re-flown to strengthen the bump bonds, then each ROC undergoes a functionality test with the sensor bias on and I-V measurements are taken. The pre-assembled HDI and base strips are glued on the accepted "raw" modules and ROCs are electrically connected to the HDI through wire bonds.

Assembled modules are moved to the testing setup which allows thermal cycling between room temperature and -20°C. The testing procedure can be divided in three steps [5]:
- DAC registers programming: all 26 DACs are set to the default value for each ROC. Most crucial DACs are tuned individually.

- Functionality test of pixel electrical connections and readout: this step includes the test of the trim bits registers and addresses as well as bump bond connections for all pixels.
- Module characterization and calibration: Noise, trims and gain response is measured for each pixel. An I-V scan is performed between 0 and 600 V. The total leakage current should not exceed 2 µA at the operational voltage of 150 V. Modules are thermally cycled between +30°C and -10°C for 24 hours and relevant parameters are measured at -10°C.

A photograph of a fully assembled FPIX panel is shown in Figure 2, right. FPIX requires 672 detector modules of five different sizes, called *plaquettes*. Each plaquette includes a silicon pixel sensor bump bonded to ROCs (from 2 to 10), glued on a two layer Very High Density Interconnect (VHDI) and supported by a 300 µm thick silicon plate. Bump bonding is performed by electroplating SnPb process at RTI (USA) and IZM (Germany) with a throughput of up to 40 plaquettes per week. Glued plaquettes are clamped in cassettes and ROCs are wire bonded to the VHDI. Before placing the plaquettes on panels thorough testing takes place, including a 2-day burn-in with temperature cycling between -15°C and +20°C, ROC characterization and measurements of bump bond yield [6].

Panels are prepared by assembling the 3-layers kapton flex circuit (HDI) on the 0.5 mm thick beryllium base plate. Passive components are soldered and the TBM chip is wire-bonded on the HDI. Finally, selected plaquettes are mounted on panels by means of vacuum transfer bars with double-sided Cho-Therm® tape. The 4 plaquette panels are mounted on the side closest to the interaction region, and the 3 plaquette versions on the opposite side.

## 4. Detector assembly and commissioning

### 4.1. *BPIX and FPIX engineering systems*

For early commissioning in the CMS experiment a BPIX *engineering system* was built including a three-layer half barrel structure with inner and outer shielding and two supply tube half cylinders. The BPIX mechanical support and supply tubes are assembled at the University of Zurich mechanical workshop. The structure is then shipped to PSI where it is equipped with 48 full modules (around 3.2 M pixels), corresponding to about 6% of the final detector.

Modules are prepared for mounting by cutting and bending the signal and power cables to the correct length with a precision of about 1–2 mm. In addition, the module position on a ladder is defined by setting the TBM chip address. Each half shell is dismounted and placed on a rotating gig. Detector modules are placed and screwed on the carbon fibre ladders using a dedicated picking tool. The estimated module mounting rate is of 32 modules per day. Printed circuit boards providing connection to the supply tube electronics for groups of 12 modules are screwed on both sides of the end flanges and module cables are connected. Each group is tested using a programmable test board replacing the Front End Driver (pxFED) and Controller (FEC) VME boards. It is foreseen to complete the assembly of the engineering structure and connection to the readout electronics by October 2007.

The pixel engineering system includes also a FPIX section. It consists of a single half cylinder equipped with two half disks. In each half disk two blades were covered with panels at FNAL, for a total of 28 plaquettes. The assembled half disks and half service cylinders were shipped to CERN where were assembled in a clean room.

During summer 2007 the FPIX engineering system was connected to the final readout electronics and power supplies. Collected data were converted in the standard CMS EDM (Root) format and stored in the CERN central tape storage system (CASTOR). Data analysis was performed with both plaquette testing software (Renaissance) and the CMS reconstruction and calibration software (CMSSW). The following tests were performed:

- Calibration runs: noise, threshold and address level separation were measured at +20°C and –10°C. The noise measured and threshold dispersion measured at -10°C was ~120 electrons and no significant difference was observed between cold and room temperature.
- Runs with radioactive source: a $^{90}$Sr source was installed behind the second half disk and particles traversing both panels of a blade were triggered with a PIN diode. Data were processed with the CMSSW reconstruction software and hits were correctly displayed by the Data Quality Monitoring software.
- Insertion test in the Silicon Strip Tracker (SST): the FPIX engineering system was inserted into the pixel support tube hosted in the SST. FPIX ground was left floating or connected to the SST ground and about a quarter of the Tracker Inner Barrel (TIB) –z side of the innermost layer was powered and triggered at 11 kHz.

A plaquette of the FPIX engineering system was pulsed and read out at ~1 kHz. Noise and threshold dispersion were measured and no significant variation was observed by changing grounding scheme or by activating the TIB power and readout.
- Test in magnetic field: the engineering system was inserted into a 4T magnetic field; calibration pulses were injected in different numbers of pixel cells. In addition, gain and noise were measured. No significant difference was observed and no broken or damaged bond wire was observed.

### 4.2. *BPIX and FPIX production detectors*

The production and testing of BPIX modules for the final detector is close to completion at PSI. Module mounting is expected to be completed in November 2007 for the first half barrel and in January 2008 for the second half. Half of the FPIX detector (–z side) is assembled and currently being commissioned at CERN. Two half disks and one service half cylinder of the +z side were recently shipped to CERN and commissioning will start in October. The shipment of the last half cylinder is expected in early November. All FED and FEC boards are available and 80% of the FEDs are installed in the CMS underground electronics room (USX55). About two thirds of the power supply boards have been delivered by CAEN (Italy) and shipment is expected to be completed by November.

The FPIX commissioning procedures at CERN start with the connection of the half cylinder to the readout electronics and power supplies. Tests can be divided in the following steps: Measurement of ROC analogue and digital current as well as sensors leakage currents; Calibration runs. The latter are of three types:
- "Pixel Alive" test: Fast run taken by pulsing every pixel cell several times. Maps of pixel response are compared with results from plaquette testing.
- Gain calibration runs: Each pixel cell is pulsed with calibration pulses of increasing amplitude and data are processed with the CMSSW software. Gain response curves have a linear range and a plateau at high pulse amplitudes. The linear range is fitted with a straight line for each pixel. The resulting offsets and gains are stored in the offline database.
- Threshold scans: the pixel response efficiency is measured for each cell by injecting calibration pulses of different amplitudes

and dividing events with non zero charge over the total number of charge injections. Runs are used to measure threshold and noise for each pixel.

Faulty or broken panels or optical transceivers are replaced and tests are repeated.

## 5. Conclusions and outlook

The FPIX and BPIX engineering systems will be installed in the CMS experiment in December 2007 and will be included in the CMS global runs with cosmic ray trigger until spring 2008. Subsequently, it will be extracted and replaced with the final and complete system. The latter is expected to be ready for installation in March 2008 and will be inserted in the CMS experiment in time for the first runs with colliding beams.

## Acknowledgements

The author thanks the CMS pixel collaboration and the organizers of the 10$^{th}$ ICATPP Conference on Astroparticle, Particle, Space Physics, Detectors and Medical Physics Applications.